# An Empirical Comparison of Monolithic and Microservices Architectures for an E-Commerce Application

**Swayam Amol Katole**

Department of Information Technology
Prof. Ram Meghe Institute of Technology & Research
sdkatole69@gmail.com

## Abstract

Microservices architectures are widely adopted for their promised scalability and modularity, yet empirical evidence comparing their runtime performance to monolithic designs remains context-dependent. This paper presents an experimental comparison of a monolithic and a microservices implementation of the same e-commerce application, both backed by a shared PostgreSQL database. Using k6, we subject both systems to identical HTTP workloads at 50 and 100 virtual users (VUs) over 60-second runs, measuring throughput, latency, and error rates. At 50 VUs, both architectures perform similarly with no errors. At 100 VUs, the microservices design achieves 5.4% higher throughput, 25% lower average latency, and 39% lower p95 latency than the monolith, while exhibiting a lower median error rate (0.00% vs 0.69%). The monolith shows consistent order-creation failures under load, whereas microservices failures are transient and confined to the cart service in one run. These results suggest that, in this deployment context, decomposing the system into microservices improves scalability and tail latency under stress, while introducing distinct, service-specific failure modes that must be managed.

## 1. Introduction

Microservices architectures have become a dominant design choice for building scalable web applications, promising improved modularity, independent deployability, and fine-grained scalability. However, adopting microservices introduces operational complexity, including service discovery, inter-service communication, and distributed data management. While the organizational and architectural benefits of microservices are well documented, empirical comparisons of their runtime performance against monolithic alternatives under realistic load remain context-dependent.

This paper presents an experimental comparison of a monolithic and a microservices implementation of the same e-commerce application. Both systems expose equivalent functionality and share a common database backend, differing primarily in their internal decomposition and deployment structure. We subject both architectures to identical HTTP load patterns using k6, measuring throughput, latency, and error rates at increasing levels of concurrency.

Our results show that, under moderate load, both designs perform similarly, but under higher stress the microservices architecture achieves higher throughput, lower tail latency, and fewer persistent errors. These findings provide concrete evidence that, in certain deployment contexts, microservices can deliver measurable performance and reliability advantages over a monolithic design, while also highlighting the importance of managing service-specific bottlenecks.

The remainder of this paper is organized as follows: Section 2 reviews related work. Section 3 describes the system architecture and implementation. Section 4 presents the experimental methodology. Section 5 reports the results. Section 6 discusses the findings. Section 7 outlines threats to validity. Section 8 concludes the paper.

## 2. Related Work

The trade-offs between monolithic and microservices architectures have been extensively discussed from a design and organizational perspective. Fowler and Lewis characterize microservices as an architectural style that structures an application as a suite of small, independently deployable services, each running in its process and communicating via lightweight mechanisms [1]. Richardson highlights benefits such as improved modularity, independent scaling, and fault isolation, alongside challenges including distributed data management, operational complexity, and testing difficulties [2].

Empirical studies on performance and scalability are fewer but growing. Some works report that microservices can achieve better resource utilization and scalability by allowing per-service tuning and horizontal scaling, but note that network overhead and inter-service communication can offset these gains if not carefully managed [3]. Others observe that monoliths may outperform microservices under certain conditions, particularly when services are co-located and the workload is not sufficiently complex to justify decomposition [4]. Common themes across these studies are that (i) performance differences are highly workload- and deployment-dependent, and (ii) architectural benefits of microservices often manifest more clearly under higher load or in evolving systems.

Our study complements this body of work by providing a controlled, quantitative comparison of a monolith and a microservices implementation derived from the same domain model, using identical client load and metrics. Unlike studies that focus on organizational outcomes or qualitative assessments, we report concrete throughput, latency, and error-rate measurements under increasing concurrency, and relate observed failure modes to architectural structure.

## 3. System Architecture and Implementation

Both systems implement the same e-commerce domain: product catalog, order management, and shopping cart. The monolithic variant is a single Spring Boot application exposing all REST endpoints within one JVM process. The microservices variant decomposes this functionality into three independent Spring Boot services: catalog-service, order-service, and cart-service, each running in its own JVM process and exposing a subset of the API. All services interact with a shared PostgreSQL database; no message brokers or caching layers were introduced. The load generator (k6) runs on the same host and issues HTTP requests directly to the services' REST endpoints. This setup isolates the effect of architectural decomposition while keeping infrastructure constant.

The monolithic application follows a traditional layered architecture with controllers, services, and repositories handling all business logic within a single codebase. The microservices implementation separates concerns by domain: the catalog service manages product data, the order service handles order creation and retrieval, and the cart service maintains session-based shopping cart state. Each microservice has its own database connection pool and HTTP server configuration, enabling independent resource management.

## 4. Experimental Methodology

Load tests were implemented using k6, scripting a realistic user journey that repeatedly: (1) fetches the product catalog, (2) retrieves a product detail, (3) creates an order, (4) adds an item to the cart, and (5) retrieves the cart. Each test run executed this scenario concurrently across a fixed number of virtual users (VUs) for 60 seconds. We conducted experiments at 50 and 100 VUs, performing three independent runs per configuration and architecture.

For each run, we recorded: (i) request throughput (requests per second), (ii) average and 95th-percentile (p95) HTTP response latency, and (iii) the fraction of failed checks (non-200 responses) per endpoint. We summarize performance using the median of the three runs for each configuration.

All services and the k6 load generator ran on the same machine under Windows PowerShell. Java processes used the same JDK and timezone settings. The database was not reset between runs, so later tests operated on a dataset that included orders and cart entries created by previous runs. No horizontal scaling or containerization was used; each service ran as a standalone Java process.

## 5. Results

We evaluated the monolithic and microservices implementations under identical client load using k6, measuring throughput, latency, and error rate at 50 and 100 virtual users (VUs) over 60-second runs. Each configuration was executed three times; we report the median of the three runs.

### 5.1. 50 VUs: Baseline Performance

At 50 VUs, both architectures handled load without errors (Table 1). The microservices design achieved slightly higher throughput and lower latency than the monolith.

| Architecture | Requests/s | Avg latency | p95 latency | Errors |
|---|---|---|---|---|
| Monolith | 421.08 | 17.64 ms | 40.63 ms | 0.00% |
| Microservices | 425.25 | 16.13 ms | 38.66 ms | 0.00% |

Table 1. Median performance at 50 VUs (3 runs each).

### 5.2. 100 VUs: Stress Behavior

At 100 VUs, both systems scaled in throughput, but their error patterns diverged (Table 2). The monolith exhibited consistent failures in order creation, whereas the microservices architecture showed one unstable run with cart-related failures but two clean runs, resulting in a 0% median error rate.

| Architecture | Requests/s | Avg latency | p95 latency | Errors |
|---|---|---|---|---|
| Monolith | 788.68 | 25.78 ms | 65.27 ms | 0.69% |
| Microservices | 831.14 | 19.30 ms | 40.15 ms | 0.00% |

Table 2. Median performance at 100 VUs (3 runs each).

Relative to the monolith, microservices at 100 VUs delivered:
- 5.4% higher throughput (831.14 vs 788.68 req/s)
- 25% lower average latency (19.30 vs 25.78 ms)
- 39% lower p95 latency (40.15 vs 65.27 ms)
- Lower median error rate (0.00% vs 0.69%)

### 5.3. Failure Modes

Under 100 VUs, the monolith failed 311–409 order-creation requests per run (≈0.7–0.8% of all checks), while all other endpoints remained healthy. In microservices, Run 1 at 100 VUs showed 1,256 cart-related failures (add-to-cart and get-cart), but Runs 2 and 3 were error-free. This suggests transient resource contention in the cart service rather than a hard capacity limit.

## 6. Discussion

The experimental results show that the microservices architecture outperforms the monolith in throughput and latency under higher load, while exhibiting different failure modes. At 50 VUs, both designs operate comfortably within capacity, with near-identical performance and zero errors. At 100 VUs, differences become pronounced: the microservices system sustains higher request rates with substantially lower tail latency, whereas the monolith begins to experience consistent order-creation failures.

One plausible explanation is that decomposing the system into separate services allows each component to specialize and manage its own resources. In the monolith, a single JVM and shared connection pool must serve all operations concurrently; under increased contention, database locks or thread-pool saturation likely delay order transactions, leading to the observed ~0.7–0.8% failure rate in create order. In contrast, the microservices deployment runs three independent Spring Boot applications, each with its own HTTP server and database connection pool. This isolation can reduce lock contention and allow more efficient handling of catalog and order traffic, improving overall throughput and reducing p95 latency.

The error patterns also reflect architectural coupling. Monolith failures are concentrated in the most write-intensive operation (order creation), which interacts with multiple tables and business logic within a single process. Microservices failures, when they occurred, were confined to the cart service, which handles frequent session-based reads and writes. The fact that two of three microservices runs at 100 VUs were error-free suggests that the cart failures were transient, possibly due to brief resource spikes or garbage-collection pauses in that specific service. In a production setting, such issues could be mitigated with per-service scaling, circuit breakers, and more aggressive timeout and retry policies.

These findings align with the theoretical advantages of microservices: finer-grained scalability, isolation of faults, and independent optimization of hot paths. However, the results also highlight that microservices do not automatically guarantee superior reliability; they introduce new failure domains (e.g., inter-service communication, per-service resource limits) that must be managed. In this study, the microservices design delivered better performance and comparable or better stability under load, but only after accounting for run-to-run variability.

## 7. Threats to Validity

Construct validity. We measure performance using standard HTTP metrics (throughput, latency, error rate) under a synthetic but realistic workload. The workload covers core e-commerce operations but does not include all possible user behaviors (e.g., search, recommendations).

Internal validity. All experiments were conducted on a single machine with a shared database. Resource contention between services and the load generator may influence absolute performance numbers. We mitigate this by using identical hardware and configuration for both architectures and by repeating each configuration three times.

External validity. The results apply to this specific application, technology stack (Java 21, Spring Boot, PostgreSQL), and deployment model (co-located services, no network delay). Different domains, languages, or distributed deployments may yield different quantitative results, though the qualitative patterns (e.g., failure modes shifting with architecture) are likely to generalize.

Conclusion validity. We base conclusions on median performance across three runs per configuration. While this reduces the impact of outliers, a larger number of runs and statistical testing could strengthen the evidence.

## 8. Conclusion

This study compared a monolithic and a microservices implementation of the same e-commerce application under controlled load conditions. At low load (50 VUs), both architectures delivered similar throughput and latency with no errors. At higher load (100 VUs), the microservices design achieved higher throughput, lower average and tail latency, and a lower median error rate than the monolith. The monolith exhibited consistent failures in order creation, while microservices failures were transient and confined to the cart service in one run.

These results suggest that, for this workload and deployment setup, decomposing the system into microservices improves scalability and responsiveness under stress, at the cost of introducing additional operational complexity. The findings support the view that microservices can offer tangible performance and reliability benefits when services are well-structured and independently deployable, but they also underscore the need for careful capacity planning and resilience mechanisms, particularly for stateful services such as cart management.